# Topological Valley Hall Edge State Lasing

Hua Zhong,[1] Yongdong Li,[1] Daohong Song,[2] Yaroslav V. Kartashov,[3] Yiqi Zhang,[1,*] Yanpeng Zhang,[1] and Zhigang Chen[2,4]

[1]*Key Laboratory for Physical Electronics and Devices of the Ministry of Education & Shaanxi Key Lab of Information Photonic Technique, Xi'an Jiaotong University, Xi'an 710049, China*

[2]*MOE Key Laboratory of Weak-Light Nonlinear Photonics, TEDA Applied Physics Institute and School of Physics, Nankai University, Tianjin 300457, China*

[3]*Institute of Spectroscopy, Russian Academy of Sciences, Troitsk, Moscow Region 108840, Russia*

[4]*Department of Physics and Astronomy, San Francisco State University, San Francisco, California 94132, USA*

*\*Corresponding author: zhangyiqi@mail.xjtu.edu.cn*

**Abstract:** Topological lasers based on topologically protected edge states offer unique features and enhanced robustness of operation in comparison with conventional lasers, even in the presence of disorder, edge deformation, and localized defects. Here we propose a new class of topological lasers arising from the valley Hall edge states, which does not require external magnetic fields or dynamical modulations of the device parameters. Specifically, topological lasing occurs at domain walls between two honeycomb waveguide arrays with broken spatial inversion symmetry. Two types of valley Hall edge lasing modes are found by shaping the gain landscape along the domain walls. In the presence of uniform losses and two-photon absorption, lasing in edge states results in the formation of stable nonlinear dissipative excitations localized on the edge of the structure, even if it has complex geometry and even if it is finite. Robustness of lasing states is demonstrated in both periodic and finite structures, where such states can circumvent sharp corners without scattering loss or radiation into the bulk. The photonic structure and mechanism proposed here for topological lasing is fundamentally different from those previously demonstrated topological lasers and can be used for fabrication of practical topological lasers of various geometries.

## 1. Introduction

Nowadays, topological photonics [1,2] is a fascinating frontier of research driven by the topology-related concepts originating from condensed matter physics [3,4]. It has attracted worldwide attention due to the unprecedented potential that topological systems bring about for manipulation of light propagation. In a photonic topological insulator, for instance, topologically protected edge states emerge as robust localized states on the edge of a bulk photonic material, immune to defects and disorders upon their evolution. Various photonic topological insulators [5-16] were realized by breaking either time-reversal symmetry [13,17] or spatial inversion symmetry [11,18-21].

Among numerous optical structures, the photonic honeycomb lattices – also called "photonic graphene" – offer a convenient platform for exploration of various topological phenomena. The spectrum of such lattices contains three pairs of degenerate but inequivalent Dirac points ($\mathbf{K}$ and $\mathbf{K}'$) at the corners of the first Brillouin zone, as has been employed for demonstration of valley pseudospin and valley Landau-Zener-Bloch oscillations [22,23]. If two sublattices of a honeycomb lattice have different refractive indices or different sizes, the inversion symmetry will be broken and a gap will open at the Dirac points, resulting in a host of fundamental new phenomena due to the intriguing valley degree of freedom [24,25]. For instance, the Berry curvature has opposite signs at the $\mathbf{K}$ and $\mathbf{K}'$ valleys, that can be attributed to the effective magnetic field leading to the well-known valley Hall effect [26]. It has been proven both theoretically and experimentally that, at the domain walls between two honeycomb lattices with inversion-symmetry breaking [27,28], there exist robust topologically protected valley Hall edge states (VHESs), so they can circumvent sharp corners without radiation into the bulk. Inspired by the discoveries in topological electronic systems, a variety of valley-mediated effects have been investigated on photonic platforms [29-32].

So far, rich physical phenomena stemming from valley degree of freedom and associated with VHESs have been considered mostly in conservative systems, leaving their dissipative counterparts largely unexplored. On the other hand, one of the most spectacular recent advances in the field of topological photonics and its technological application is the realization of topological lasers [33-38], which are essentially dissipative systems. In such laser systems, lasing occurs based on topologically protected edge states and therefore exhibits features outperforming conventional lasers whose stability may be affected by perturbations such as defects and disorder. In addition to lasing based on edge states in a one-dimensional Su-Schrieffer-Heeger chain [33-36], two-dimensional topological lasing has been achieved in photonic crystals [39] and lattices of coupled-ring resonators [40,41], and has been proposed theoretically for polaritonic arrays [42], Floquet topological insulators [43] and a bosonic Harper-Hofstadter model [44]. However, topological lasers based on VHESs that do not require external magnetic fields or dynamical modulations of system parameters have not been explored to the best of our knowledge.

The aim of this work is to introduce two-dimensional VHES lasers that can be implemented using photonic honeycomb waveguides arrays with broken spatial inversion symmetry fabricated in conventional nonlinear optical transparent materials with gain saturation. We show that topologically protected VHESs at the domain wall in this system can lase if spatially inhomogeneous gain is provided. Such topological VHES lasers do not require magnetic fields for their operation and do not rely on judicious engineering of coupling between elements as in coupled-resonator arrays. In addition, in comparison with helical waveguide arrays on which the majority of previous topological photonic systems were constructed, straight waveguide arrays taken here are more feasible for experimental realization and are free from radiative losses typical for Floquet systems.

## 2. The model and spectrum of the system

The propagation dynamics of light beams in our dissipative structure admitting VHESs can be described by the nonlinear Schrödinger-like equation that in dimensionless units reads as:

$$i\frac{\partial \psi}{\partial z}=-\frac{1}{2}\nabla^2\psi-[\mathcal{R}_{\mathrm{re}}-i\mathcal{R}_{\mathrm{im}}+i\gamma]\psi-(1+i\alpha)|\psi|^2\psi. \quad (1)$$

Here, $\psi=(\kappa^2 w^2 n_{2,\mathrm{re}}/n_{\mathrm{re}})^{1/2}E$ is the scaled field amplitude; $x,y$ are the transverse coordinates normalized to the characteristic scale $w$; $z$ is the propagation distance scaled to the diffraction length $\kappa w^2$; $\kappa=2\pi n_{\mathrm{re}}/\lambda$ is the wavenumber; $n_{\mathrm{re}}$ and $n_{\mathrm{im}}$ ($n_{\mathrm{im}} \ll n_{\mathrm{re}}$) are the real and imaginary parts of the unperturbed linear refractive index of the material, respectively; $n_{2,\mathrm{re}}$ and $n_{2,\mathrm{im}}$ are the real and imaginary parts of the nonlinear refractive index, respectively; $\gamma=\kappa^2 w^2 n_{\mathrm{im}}/n_{\mathrm{re}}$ is the coefficient of linear losses that are assumed uniform; $\alpha=n_{2,\mathrm{im}}/n_{2,\mathrm{re}}$ is the scaled coefficient characterizing nonlinear losses stemming from all sources, including intrinsic nonlinear losses of the medium and gain saturation in the first approximation. Further we consider focusing cubic (Kerr) nonlinearity, typical for many solid materials, including optical fibers. We assume that the VHES laser is composed of two honeycomb arrays but with broken inversion symmetry. The refractive index distribution in each array is described by the function $\mathcal{R}_{\mathrm{re}}(x,y)=\mathcal{R}_{\mathrm{re}}^{\mathrm{A}}(x,y)+\mathcal{R}_{\mathrm{re}}^{\mathrm{B}}(x,y)$, where $\mathcal{R}_{\mathrm{re}}^{\mathrm{A}}$ and $\mathcal{R}_{\mathrm{re}}^{\mathrm{B}}$ stand for two standard sublattices of the honeycomb array. Each sublattice $\mathcal{R}_{\mathrm{re}}^{\mathrm{A,B}}(x,y)=p_{\mathrm{re}}^{\mathrm{A,B}}\sum_{n,m}\mathcal{Q}(x-x_n,y-y_m)$ is composed of Gaussian waveguides $\mathcal{Q}=\exp[-(x^2+y^2)/d^2]$ with normalized depths of $p_{\mathrm{re}}^{\mathrm{A,B}}=\kappa^2 w^2 \delta n_{\mathrm{re}}^{\mathrm{A,B}}/n_{\mathrm{re}}$, where $(x_n,y_m)$ are the coordinates of the sites of the honeycomb lattice and $d$ the waveguide width. The separation between the waveguides in the array is denoted as $b$. Further we introduce detuning between two sublattices, i.e. we set $p_{\mathrm{re}}^{\mathrm{A}} > p_{\mathrm{re}}^{\mathrm{B}}$ (further we choose $p_{\mathrm{re}}^{\mathrm{A}}=7$ and $p_{\mathrm{re}}^{\mathrm{B}}=6$, and we assume that every channel is a single-mode waveguide). An armchair domain wall [20,32,45-53] is created at the interface between such an array and another honeycomb array with inverted detuning $p_{\mathrm{re}}^{\mathrm{A}} < p_{\mathrm{re}}^{\mathrm{B}}$ between sublattices, as shown in Fig. 1(a) by the dashed rectangle. Gain is only provided on the armchair domain wall, but selectively on the pairs of identical sites with either deeper or shallower potential, as shown in Figs. 1(b,f). The gain is also described by the function $\mathcal{R}_{\mathrm{im}}(x,y,z)=p_{\mathrm{im}}\sum_{q,l}\mathcal{Q}(x-x_q,y-y_l)$, where $x_q,y_l$ are the coordinates of domain wall waveguides, and $p_{\mathrm{im}}=\kappa^2 w^2 \delta n_{\mathrm{im}}/n_{\mathrm{re}}$ is the gain amplitude ($p_{\mathrm{im}} \ll p_{\mathrm{re}}^{\mathrm{A,B}}$). For convenience, we label the gain profile in Fig. 1(b) as type-I (gain on deeper sites), and that in Fig. 1(f) as type-II (gain on shallower sites). In the following, we will explain why we consider these two types of gain landscapes. The period of this structure in the $y$-direction equals to $Y=3^{1/2}b$.

We would like to mention the possibility of implementation of this scheme in pumped planar periodic structures, like those recently used in [39,54]. In the particular case of doped chalcogenide glasses, such as GaLaS or AsSe with nonlinear index $n_{2,\mathrm{re}} \sim 1\times10^{-17}\,\mathrm{m}^2/\mathrm{W}$ and absorption coefficient ranging from $n_{2,\mathrm{im}} \sim 2\times10^{-19}\,\mathrm{m}^2/\mathrm{W}$ to $n_{2,\mathrm{im}} \sim 1\times10^{-17}\,\mathrm{m}^2/\mathrm{W}$ depending on composition [55-57], for the characteristic transverse scale of $w \sim 10\,\mu\mathrm{m}$ and unperturbed refractive index $n_{\mathrm{re}} \sim 2.81$ at the wavelength of $\lambda=1.08\,\mu\mathrm{m}$ one finds that the dimensionless diffraction length is $\sim 1.5\mathrm{mm}$; the refractive index modulation depth of $p_{\mathrm{re}}^{\mathrm{A,B}}=6$ corresponds to real refractive index modulation depth $\sim 6.8\times10^{-4}$; while parameter $p_{\mathrm{im}}=0.1$ corresponds to $\delta n_{\mathrm{im}} \sim 1\times10^{-5}$.

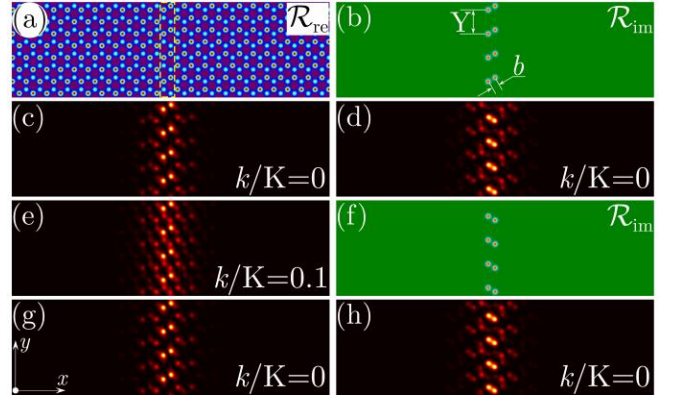

Fig. 1. (a) Real part of the refractive index in the array with an armchair domain wall (indicated by a dashed rectangle). (b) Gain landscape of type-I with amplification on deeper sites of the armchair domain wall. (c) Intensity profile of the VHES with $k=0$ that has intensity maxima on deeper sites. (d) Intensity profile of the VHES with $k=0$ that has intensity maxima on shallower sites. (e) Same as (c), but for $k=0.1\mathrm{K}$. (f) Gain landscape of type-II with amplification on shallower sites of the armchair domain wall and (g,h) corresponding edge states of two different types. Other parameters are $d=0.4$, $b=1.4$, and $p_{\mathrm{im}}=0.35$.

As a first step, we consider linear modes of our structure by setting $\alpha=0$ and neglecting the focusing nonlinearity, but keeping the linear loss $\gamma=0.05$ and inhomogeneous gain $\mathcal{R}_{\mathrm{im}}$. We seek for the eigenmodes of the form $\psi(x,y,z)=u(x,y)\exp(i\epsilon z+iky)$ in the corresponding linear equation, where $u(x,y)=u(x,y+Y)$, $k$ is the Bloch momentum along the $y$ direction, and $\epsilon=\epsilon_{\mathrm{re}}+i\epsilon_{\mathrm{im}}$ is the "energy" with $\epsilon_{\mathrm{re}}$ and $\epsilon_{\mathrm{im}}$ being the real and imaginary parts, respectively. The sign of the imaginary part $\epsilon_{\mathrm{im}}$ is determined by the loss $\gamma$ and gain $\mathcal{R}_{\mathrm{im}}$. If $\epsilon_{\mathrm{im}} < 0$, the modes are amplified, while if $\epsilon_{\mathrm{im}} > 0$ the modes are damped. Figure 2 shows $\epsilon_{\mathrm{re}}$ and $\epsilon_{\mathrm{im}}$ versus normalized Bloch momentum $k/\mathrm{K}$ with $\mathrm{K}=2\pi/Y$ for different gain amplitudes $p_{\mathrm{im}}$. In the band gap, there are two VHESs throughout the first Brillouin zone, which are indicated by the red and blue colors. The red edge state is concentrated mainly on deeper sites of the domain wall [Fig. 1(c)], while blue edge state resides mainly on shallower sites [Fig. 1(d)]. This opens unique opportunity to selectively amplify necessary state by providing gain either on deeper or on shallower sites (since overall amplification exhibited by the edge state is determined by the overlap integral of its profile and gain landscape) - the reason why we introduced type-I and type-II gain landscapes. In Figs. 2(a,b) linear band structures are plotted for the type-I gain landscape with $p_{\mathrm{im}}=0.3$ and $p_{\mathrm{im}}=0.4$. One can see that the imaginary part of the energy may assume negative values on edge state branches around $k=0$ (in the figure we plot the inverted value $-\epsilon_{\mathrm{im}}$ for illustrative purposes without showing the $-\epsilon_{\mathrm{im}} < 0$ region corresponding to damped modes, hence color spikes in Fig. 2 directly show which modes will grow upon evolution). Numerical simulations demonstrate that this happens if gain amplitude $p_{\mathrm{im}}$ exceeds a threshold $p_{\mathrm{im}}^{\mathrm{th}} \sim 0.282$ defining threshold for lasing in VHESs. Comparing Figs. 2(a,b), we find that the Bloch momentum interval centered on $k=0$, where edge states are amplified, expands with increasing gain amplitude $p_{\text{im}}$. For large gain levels $p_{\text{im}}$ lasing occurs in the entire first Brillouin zone for red edge state (that always feature largest effective gain for type-I landscape), and even for bulk and blue edge states. We display red edge states obtained at $p_{\text{im}} = 0.35$ at different momenta in Figs. 1(c,e), while blue edge state at $k=0$ is shown in Fig. 1(d). The red state in Fig. 1(c) is amplified, but blue state in Fig. 1(d) is damped. If the momentum of the red edge state is located out of the range, where $-\epsilon_{\text{im}} > 0$, this state is damped as well. An example is shown in Fig. 1(e) that is out of the lasing range $-0.065\text{K} \le k \le 0.065\text{K}$ for $p_{\text{im}} = 0.35$. This picture inverts for the type-II gain profile, see corresponding band structures in Figs. 2(c,d). Now one finds that the blue edge states features the lowest lasing threshold, which is about $p_{\text{im}}^{\text{th}} \sim 0.308$ and is higher than the lasing threshold for red edge state. The red and blue VHESs at $k=0$ obtained for type-II gain landscape for parameters of Fig. 2(c) are displayed in Figs. 1(g,h), respectively, but this time the former state is attenuated and the latter one is amplified. Providing gain on both deep and shallow sites of the armchair domain wall may lead to simultaneous lasing in red and blue edge states, but this is supposed to lead to beatings that should be avoided if single-mode operation is desired. In what follows, without loss of generality we will consider type-I gain landscape, unless stated otherwise.

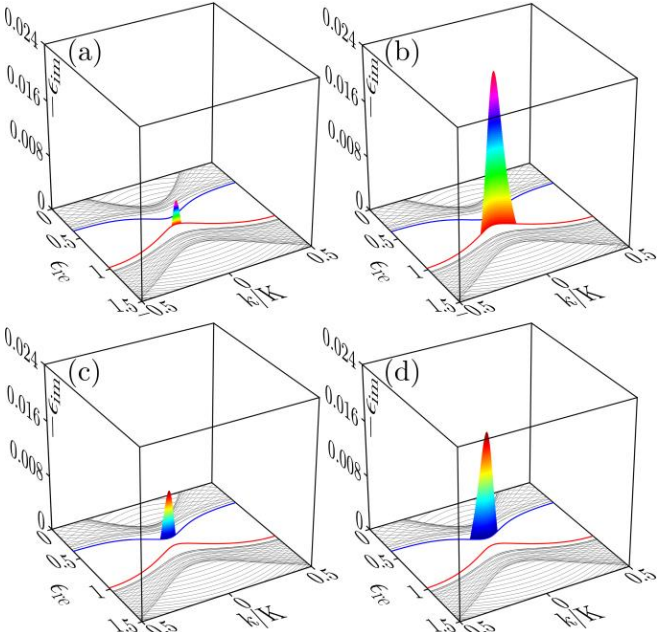

Fig. 2. Real $\epsilon_{\text{re}}$ and imaginary $\epsilon_{\text{im}}$ parts of energy of the linear VHESs. (Top) Type-I gain landscape on the armchair domain wall, (a) $p_{\text{im}} = 0.3$ and (b) $p_{\text{im}} = 0.4$. (Bottom) Type-II gain landscape, (c) $p_{\text{im}} = 0.35$ and (d) $p_{\text{im}} = 0.4$. Only states with $-\epsilon_{\text{im}} > 0$ can lase, while all other states are attenuated.

## 3. Valley Hall lasing state

Amplification of the edge states at $p_{\text{im}} > p_{\text{im}}^{\text{th}}$ [see corresponding dependence of the imaginary and real part of the energy of the amplified edge state at $k=0$ on $p_{\text{im}}$ illustrated in Fig. 3(a), where lasing domain associated with $-\epsilon_{\text{im}} > 0$ lies below the blue plane in the figure] can be eventually arrested by the nonlinear absorption. To explore the possibility of the exact and stable balance between diffraction, nonlinearity, gain and absorption in this system we now consider model (1) with all nonlinear terms included and search for stationary nonlinear edge states with constant power along propagation distance. Since several VHESs may experience amplification above the lasing threshold, we use linear mode of the conservative system with particular $k$ value as an input for nonlinear Eq. (1) and solve it up to sufficiently large distance. In subsequent evolution, the competition between modes typically results in the emergence of stationary nonlinear dissipative edge state that we further use as an input to build the entire family of dissipative solutions parameterized by gain amplitude $p_{\text{im}}$. The obtained solutions at selected value of the momentum $k$ were additionally checked by the Newton method complemented by the power balance condition.

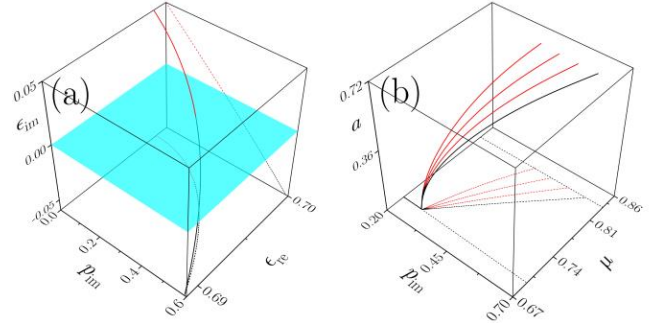

Fig. 3. (a) Real $\epsilon_{\text{re}}$ and imaginary $\epsilon_{\text{im}}$ parts of energy of the linear edge state with $k=0$ versus $p_{\text{im}}$. Around $p_{\text{im}} \sim 0.282$, when the curve crosses blue plane, lasing in VHES occurs. (b) Amplitude $a$ and nonlinear energy shift $\mu$ of the edge state with $k=0$ versus $p_{\text{im}}$ for different nonlinear absorption coefficients increasing from $\alpha = 0.2$ to $\alpha = 0.5$ in steps of $0.1$. Lower dashed line in $(p_{\text{im}}, \mu)$ plane indicates the energy of the linear edge state; upper dashed line is the boundary of the band gap (viz. bottom of bulk band). Stable branches are shown black and unstable branches are shown red.

To characterize nonlinear families of the VHESs in our laser we determine their peak amplitude $a = \max|\psi|$ and the nonlinear energy shift (or propagation constant) $\mu$, since in stationary states $\psi \sim \exp(i\mu z)$. Figure 3(b) show the dependencies of $a$ and $\mu$ on gain amplitude $p_{\text{im}}$ for $k=0$ at different values of the nonlinear absorption coefficient $\alpha$. In Fig. 3(b) the lasing threshold in $p_{\text{im}}$ is clearly seen, where the amplitude of the lasing state becomes nonzero. Lasing in the states with nonzero momentum $k$ can occur too, but the threshold for lasing will be higher in this case. This is because the state with $k=0$ has largest overlap with gain landscape leading to the most efficient amplification. Energy shift $\mu$ below the lasing threshold naturally coincides with the energy of the linear edge state (corresponding state is damped), as shown by the bottom dotted line in the $(p_{\text{im}}, \mu)$ plane in Fig. 3(b). Above the threshold, nonlinear energy shift increases almost linearly with gain amplitude until it reaches the bottom of the bulk band, as indicated by the upper dotted line in the $(p_{\text{im}}, \mu)$ plane in Fig. 3(b). Since we only consider truly localized lasing states formed in the band gap, we truncate the curves in Fig. 3(b) accordingly. By comparing the curves in Fig. 3(b) for different nonlinear absorption coefficients, one finds that the interval of gain amplitudes where localized VHESs exist expands with the growth of the nonlinear absorption $\alpha$.

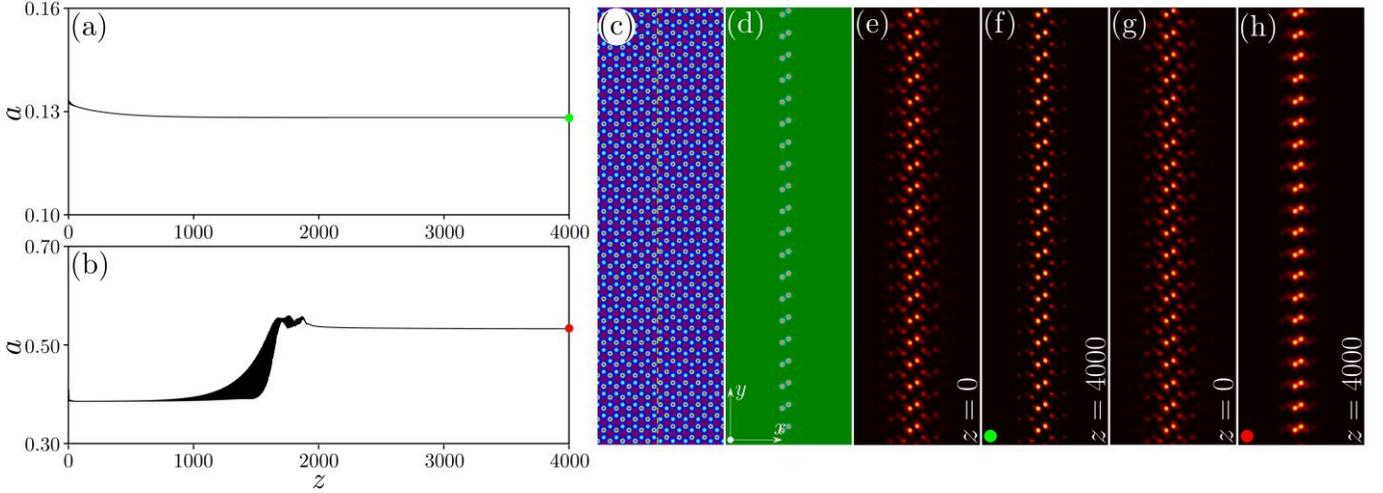

Fig. 4. Peak amplitude $a = \max|\psi|$ versus distance $z$ illustrating (a) stable propagation of the lasing state at $p_{\rm im} = 0.29$, $\alpha = 0.4$, $k = 0$ [$|\psi|^2$ distributions at $z = 0$ and $z = 4000$ are shown in (e,f)], and (b) unstable propagation at $p_{\rm im} = 0.36$, $\alpha = 0.4$, $k = 0$ [corresponding $|\psi|^2$ distributions at $z = 0$ and $z = 4000$ are shown in (g,h)]. (c) Refractive index distribution with a domain wall indicated by a dashed line. (d) Type-I gain profile.

Exemplary stable and unstable lasing states for $k=0$ are shown in Fig. 4 together with the array profile and gain landscape [Figs. 4(c,d)]. In Fig. 4(a), the peak amplitude $a = \max|\psi|$ of the perturbed VHES at $p_{\rm im} = 0.29$ and $\alpha = 0.4$ is displayed as a function of the propagation distance. The peak amplitude quickly returns to the unperturbed value and remains unchanged up to the propagation distance $z > 4000$, demonstrating stability of the lasing state. We also show typical intensity distributions in the lasing state at $z = 0$ and $z = 4000$ [green dot in Fig. 4(a)] in Figs. 4(e,f), which are almost the same except for a small difference due to the artificially introduced perturbation in Fig. 4(e). In contrast, unstable evolution at $p_{\rm im} = 0.36$ is illustrated in Fig. 4(b). Interestingly, after the transient stage $1000 < z < 2000$, where the instability develops, the wave transforms into different stable pattern with invariable peak amplitude, see intensity distribution at $z = 4000$ in Fig. 4(g) [corresponding to the red dot in Fig. 4(b)]. The nonlinear energy shift for this state is $\mu \sim 1.57$ that falls into the allowed band, rather than into gap. This indicates on the presence of the small-amplitude background in this mode in the bulk of the array, which however is practically invisible on the scale of Fig. 4(g). This mode also has phase distribution that is substantially different from that in nonlinear state residing in the gap (neighboring spots on deep channels are not out-of-phase anymore).

Since practical laser system should be spatially compact, here we propose the design of such spatially limited structures. Two different honeycomb lattices can be placed in contact such that domain wall exhibits relatively sharp bends allowing to create closed-contour configurations. Such structures with finite length of the edge are supposed to be beneficial for stability of edge modes, since they eliminate destructive perturbations with very large periods exceeding the length of the edge. Here, we design domains with triangular, hexagonal and rhombus shapes [Figs. 5(a1-c1)] and type-I gain profile in all the cases [Figs. 5(a2-c2)].

If a moving linear VHES corresponding to $k = 0.1\mathrm{K}$ with a broad Gaussian envelope $\exp(-y^2/w^2)$ of the width $w = 9b$ is launched into such structures on their left vertical edges, one observes gradual transition from circulation of this localized state along the perimeter of the structure to lasing on the entire perimeter. This transition is illustrated with isosurfaces plots and insets showing intensity distributions at different distances in triangular [Fig. 6(a)], rhombic [Fig. 6(b)] and hexagonal [Fig. 6(c)] finite structures. Notice that for selected parameters one observes at least two complete roundtrips of the localized initial state over the perimeter of the structure without noticeable reflections in the corners before amplification counterbalanced by the nonlinear absorption leads to transition to stable lasing along the entire perimeter (this typically happens at distances $z \sim 1500$). Notice the absence of radiation in the bulk of this structure illustrating topological regime of operation.

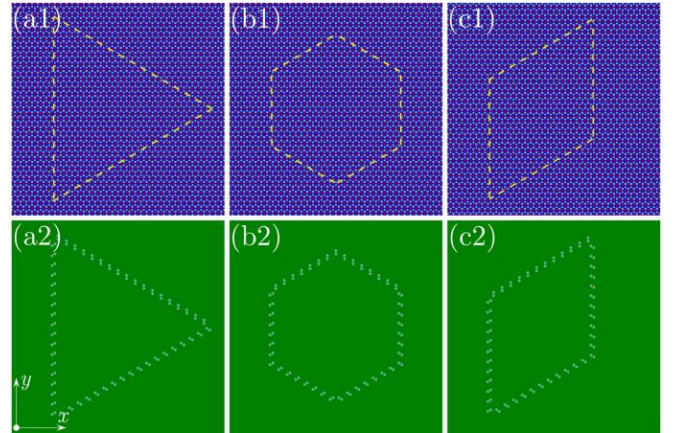

Fig. 5. Practical laser systems with different geometries. (a1,a2) Triangular structure. (b1,b2) Hexagonal structure. (c1,c2) Rhombic structure. Top row: refractive index distributions with dashed lines indicating the domain wall. Bottom row: type-I gain landscape.

## 4. Summary

In conclusion, we have proposed the VHES lasers in inversion symmetry broken honeycomb lattices with gain applied on the armchair domain wall. The balance between diffraction, focusing nonlinearity, uniform loss, nonlinear absorption, and gain, may allow lasing and formation of topologically protected nonlinear VHESs. We found that increasing gain could destabilize the lasing states, while increasing nonlinear absorption broadens their stability intervals. Compact VHES lasers with different geometries were also designed. This work paves the way to realization of topological lasers without using magnetic fields and may inspire the investigation of topological VHESs in non-Hermitian systems.

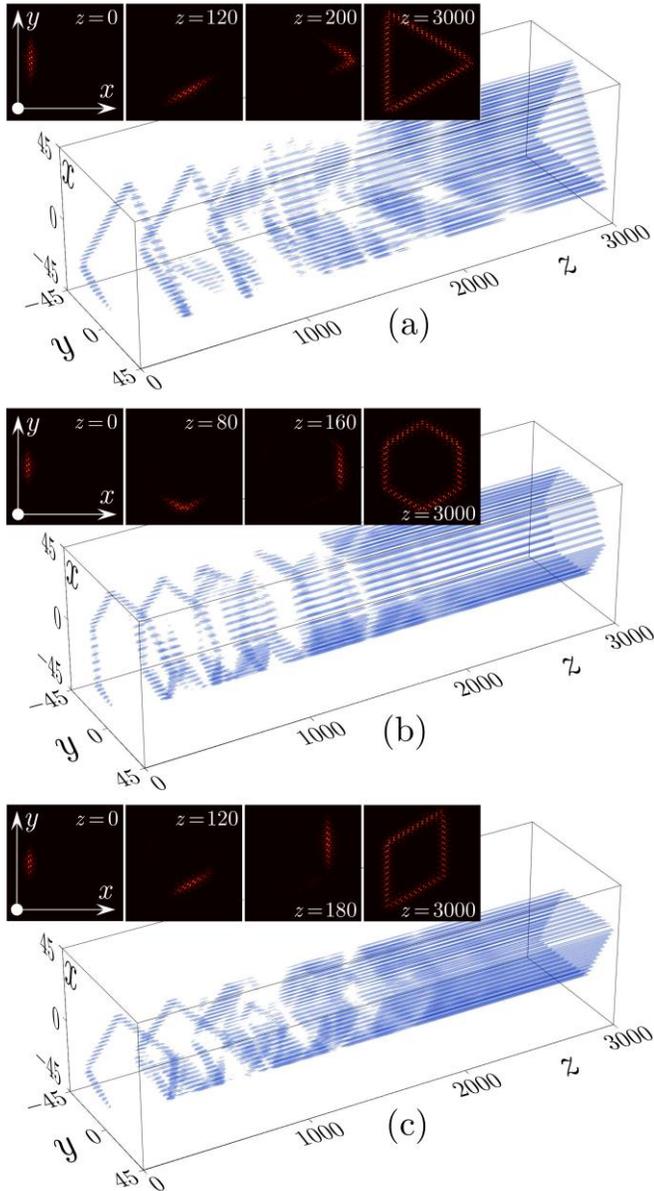

Fig. 6. Topological laser patterns of moving VHESs with different geometries. Circulation of the VHESs and transition to stable lasing in triangular (a), hexagonal (b), and rhombic (c) structures at $\gamma = 0.005$, $p_{\mathrm{im}} = 0.045$ and $\alpha = 0.3$. The initial state corresponds to the linear edge state with momentum $k = 0.1\mathrm{K}$ with superimposed envelope of the width $w = 9b$.

## Acknowledgments

Y.V.K. acknowledges funding of this study by RFBR and DFG according to the research project No. 18-502-12080. This research is supported by the National Key R&D Program of China (2017YFA0303800), the Natural Science Foundation of Guangdong Province (2018A0303130057), and the Fundamental Research Funds for the Central Universities (xzy012019038 and xzy022019076). We acknowledge Dr. Baile Zhang for helpful discussions and the computational resources provided by the HPC platform of Xi'an Jiaotong University.

## Conflict of Interest

The authors declare no conflict of interest.

## Keywords

Topological Laser; Valley Hall Effect; Topological Insulator; Edge state.

Received: ((will be filled in by the editorial staff))
Revised: ((will be filled in by the editorial staff))
Published online: ((will be filled in by the editorial staff))

## References

[1] L. Lu, J. D. Joannopoulos, and M. Soljacic, Topological photonics, Nat. Photon. **8**, 821–829 (2014).

[2] T. Ozawa, H. M. Price, A. Amo, N. Goldman, M. Hafezi, L. Lu, M. Rechtsman, D. Schuster, J. Simon, O. Zilberberg, and I. Carusotto, "Topological photonics," Rev. Mod. Phys. **91**, 015006 (2019).

[3] M. Z. Hasan and C. L. Kane, "Colloquium: Topological insulators," Rev. Mod. Phys. **82**, 3045-3067 (2010).

[4] X.-L. Qi and S.-C. Zhang, "Topological insulators and superconductors," Rev. Mod. Phys. **83**, 1057 (2011).

[5] F. D. M. Haldane and S. Raghu, "Possible Realization of Directional Optical Waveguides in Photonic Crystals with Broken Time-Reversal Symmetry," Phys. Rev. Lett. **100**, 013904 (2008).

[6] Z. Wang, Y. Chong, J. D. Joannopoulos, and M. Soljacic, "Observation of unidirectional backscattering-immune topological electromagnetic states," Nature **461**, 772-775 (2009).

[7] N. H. Lindner, G. Refael, and V. Galitski, "Floquet topological insulator in semiconductor quantum wells," Nat. Phys. **7**, 490-495 (2011).

[8] M. Hafezi, E. A. Demler, M. D. Lukin, and J. M. Taylor, "Robust optical delay lines with topological protection," Nat. Phys. **7**, 907-912 (2011).

[9] R. O. Umucalilar and I. Carusotto, "Fractional Quantum Hall States of Photons in an Array of Dissipative Coupled Cavities," Phys. Rev. Lett. **108**, 206809 (2012).

[10] M. Hafezi, S. Mittal, J. Fan, A. Migdall, and J. M. Taylor, "Imaging topological edge states in silicon photonics," Nat. Photon. **7**, 1001-1005 (2013).

[11] A. B. Khanikaev, S. H. Mousavi, W.-K. Tse, M. Kargarian, A. H. MacDonald, and G. Shvets, "Photonic topological insulators," Nat. Mater. **12**, 233-239 (2013).

[12] W.-J. Chen, S.-J. Jiang, X.-D. Chen, B. Zhu, L. Zhou, J.-W. Dong, and C. T. Chan, "Experimental realization of photonic topological insulator in a uniaxial metacrystal waveguide," Nat. Commun. **5**, 5782 (2014).

[13] M. C. Rechtsman, J. M. Zeuner, Y. Plotnik, Y. Lumer, D. Podolsky, F. Dreisow, S. Nolte, M. Segev, and A. Szameit, "Photonic Floquet topological insulators," Nature **496**, 196-200 (2013).

[14] S. Stützer, Y. Plotnik, Y. Lumer, P. Titum, N. H. Lindner, M. Segev, M. C. Rechtsman, and A. Szameit, "Photonic topological Anderson insulators," Nature **560**, 461-465 (2018).

[15] Y. Yang, Z. Gao, H. Xue, L. Zhang, M. He, Z. Yang, R. Singh, Y. Chong, B. Zhang, and H. Chen, "Realization of a three-dimensional photonic topological insulator," Nature **565**, 622-626 (2019).

[16] E. Lustig, S. Weimann, Y. Plotnik, Y. Lumer, M. A. Bandres, A. Szameit, and M. Segev, "Photonic topological insulator in synthetic dimensions," Nature **567**, 356-360 (2019).

[17] S. Klembt, T. H. Harder, O. A. Egorov, K. Winkler, R. Ge, M. A. Bandres, M. Emmerling, L. Worschech, T. C. H. Liew, M. Segev, C. Schneider, and S. Höfling, "Exciton-polariton topological insulator," Nature **562**, 552-556 (2018).

[18] L.-H. Wu and X. Hu, "Scheme for Achieving a Topological Photonic Crystal by Using Dielectric Material," Phys. Rev. Lett. **114**, 223901 (2015).

[19] X. Wu, Y. Meng, J. Tian, Y. Huang, H. Xiang, D. Han, and W. Wen, "Direct observation of valley-polarized topological edge states in designer surface plasmon crystals," Nat. Commun. **8**, 1304 (2017).

[20] J. Noh, S. Huang, K. P. Chen, and M. C. Rechtsman, Observation of photonic topological valley Hall edge states, Phys. Rev. Lett. **120**, 063902 (2018).

[21] H. Zhong, Y. V. Kartashov, Y. Q. Zhang, D. H. Song, Y. P. Zhang, F. L. Li, and Z. Chen, "Rabi-like oscillation of photonic topological valley Hall edge states," Opt. Lett. **44**, 3342-3345(2019).

[22] D. Song, D. Leykam, J. Su, X. Liu, L. Tang, S. Liu, J. Zhao, N. K. Efremidis, J. Xu, and Z. Chen, "Valley vortex states and degeneracy lifting via photonic higher-band excitation," Phys. Rev. Lett. **122**, 123903 (2019).

[23] Y. Sun, D. Leykam, S. Nenni, D. Song, H. Chen, Y. D. Chong, and Z. Chen, "Observation of Valley Landau-Zener-Bloch Oscillations and Pseudospin Imbalance in Photonic Graphene," Phys. Rev. Lett. **121**, 033904 (2018).

[24] D. Xiao, W. Yao, and Q. Niu, "Valley-Contrasting Physics in Graphene: Magnetic Moment and Topological Transport," Phys. Rev. Lett. **99**, 236809 (2007).

[25] W. Yao, D. Xiao, and Q. Niu, "Valley-dependent optoelectronics from inversion symmetry breaking," Phys. Rev. B **77**, 235406 (2008).

[26] K. F. Mak, K. L. McGill, J. Park, and P. L. McEuen, "The valley Hall effect in $MoS_2$ transistors," Science **344**, 1489–1492 (2014).

[27] O. Bleu, G. Malpuech, and D. D. Solnyshkov, "Robust quantum valley Hall effect for vortices in an interacting bosonic quantum fluid," Nat. Commun. **9**, 3991 (2018).

[28] X.-D. Chen, F.-L. Zhao, M. Chen, and J.-W. Dong, "Valley-contrasting physics in all-dielectric photonic crystals: Orbital angular momentum and topological propagation," Phys. Rev. B **96**, 020202 (2017).

[29] J.-W. Dong, X.-D. Chen, H. Zhu, Y. Wang, and X. Zhang, "Valley photonic crystals for control of spin and topology," Nat. Mater. **16**, 298-302 (2017).

[30] X.-T. He, E.-T. Liang, J.-J. Yuan, H.-Y. Qiu, X.-D. Chen, F.-L. Zhao, and J.-W. Dong, "A silicon-on-insulator slab for topological valley transport," Nat. Commun. **10**, 872 (2019).

[31] F. Deng, Y. Sun, X. Wang, R. Xue, Y. Li, H. Jiang, Y. Shi, K. Chang, and H. Chen, "Observation of valley-dependent beams in photonic graphene," Opt. Express **22**, 23605–23613 (2014).

[32] F. Gao, H. Xue, Z. Yang, K. Lai, Y. Yu, X. Lin, Y. Chong, G. Shvets, and B. Zhang, "Topologically protected refraction of robust kink states in valley photonic crystals," Nat. Phys. **14**, 140-144 (2017).

[33] L. Pilozzi and C. Conti, Topological lasing in resonant photonic structures, Phys. Rev. B **93**, 195317 (2016).

[34] P. St-Jean, V. Goblot, E. Galopin, A. Lemaítre, T. Ozawa, L. Le Gratiet, I. Sagnes, J. Bloch, and A. Amo, "Lasing in topological edge states of a 1D lattice," Nat. Photon. **11**, 651-656 (2017).

[35] M. Parto, S. Wittek, H. Hodaei, G. Harari, M. A. Bandres, J. Ren, M. C. Rechtsman, M. Segev, D. N. Christodoulides, and M. Khajavikhan, "Edge-Mode Lasing in 1D Topological Active Arrays," Phys. Rev. Lett. **120**, 113901 (2018).

[36] H. Zhao, P. Miao, M. H. Teimourpour, S. Malzard, R. El-Ganainy, H. Schomerus, and L. Feng, Topological hybrid silicon microlasers, Nat. Commun. **9**, 981 (2018).

[37] S. Longhi, Y. Kominis, and V. Kovanis, "Presence of temporal dynamical instabilities in topological insulator lasers," EPL **122**, 14004 (2018).

[38] S. Malzard and H. Schomerus, "Nonlinear mode competition and symmetry-protected power oscillations in topological lasers," New J. Phys. **20**, 063044 (2018).

[39] B. Bahari, A. Ndao, F. Vallini, A. El Amili, Y. Fainman, and B. Kanté, "Nonreciprocal lasing in topological cavities of arbitrary geometries," Science **358**, 636 (2017).

[40] G. Harari, M. A. Bandres, Y. Lumer, M. C. Rechtsman, Y. D. Chong, M. Khajavikhan, D. N. Christodoulides, and M. Segev, "Topological insulator laser: Theory," Science **359**, eaar4003 (2018).

[b41] M. A. Bandres, S. Wittek, G. Harari, M. Parto, J. Ren, M. Segev, D. N. Christodoulides, and M. Khajavikhan, "Topological insulator laser: Experiment," Science **359**, eaar4005 (2018).

[42] Y. V. Kartashov and D. V. Skryabin, "Two-Dimensional Topological Polariton Laser," Phys. Rev. Lett. **122**, 083902 (2019).

[43] S. K. Ivanov, Y. Q. Zhang, Y. V. Kartashov, and D. V. Skryabin, "Floquet topological insulator laser," APL Photon. **4**, 126101(2019).

[44] M. Seclì, M. Capone, and I. Carusotto, "Theory of chiral edge state lasing in a two-dimensional topological system," Phys. Rev. Research **1**, 033148 (2019).

[45] A. Drouot and M. I. Weinstein, "Edge states and the valley Hall effect," arXiv:1910.03509.

[46] J. Li, R.-X. Zhang, Z. Yin, J. Zhang1, K. Watanabe, T. Taniguchi, C. Liu, J. Zhu, "A valley valve and electron beam splitter," Science **362**, 1149–1152 (2018).

[47] L. Zhang, Y. Yang, M. He, H.-X. Wang, Z. Yang, E. Li, F. Gao, B. Zhang, R. Singh, J.-H. Jiang, and H. Chen, "Valley Kink States and Topological Channel Intersections in Substrate-Integrated Photonic Circuitry," Laser Photon. Rev. **13**, 1900159 (2019).

[48] T. Ma and G. Shvets, "All-Si valley-Hall photonic topological insulator," New J. Phys. **18**, 025012 (2016).
[49] Z. Qiao, J. Jung, Q. Niu, and A. H. MacDonald, "Electronic Highways in Bilayer Graphene," Nano Lett. **11**, 3453–3459 (2011).
[50] Z. Qiao, J. Jung, C. Lin, Y. Ren, A. H. MacDonald, and Q. Niu, "Current Partition at Topological Channel Intersections," Phys. Rev. Lett. **112**, 206601 (2014).
[51] Y. Ren, Z. Qiao, and Q. Niu, "Topological phases in two-dimensional materials: a review," Rep. Prog. Phys. **79**, 066501 (2016).
[52] S. Yves, R. Fleury, T. Berthelot, M. Fink, F. Lemoult, and G. Lerosey, "Crystalline metamaterials for topological properties at subwavelength scales," Nat. Commun. **8**, 16023 (2016).
[53] A. B. Khanikaev and G. Shvets, "Two-dimensional topological photonics," Nat. Photon. **11**, 763–773 (2017).
[54] Z.-K. Shao, H.-Z. Chen, S. Wang, X.-R. Mao, Z.-Q. Yang, S.-L. Wang, X.-X. Wang, X. Hu, and R.-M. Ma, "A high-performance topological bulk laser based on band-inversion-induced reflection," Nat. Nanotechnol. (2019) doi: 10.1038/s41565-019-0584-x
[55] A. Zakery and S. R. Elliott, *Optical nonlinearities in chalcogenide glasses and their applications* (Springer, Berlin, 2007).
[56] F. Smektala, C. Quemard, V. Couderc, and A. Barthelemy, "Non-linear optical properties of chalcogenide glasses measured by Z-scan," J. Non-Crystalline Solids **274**, 232-237 (2000).
[57] J. Harrington, *Infrared Fibers and Their Applications*, (SPIE Press, Bellingham, WA, 2003).